\documentstyle[11pt]{article}
\def\one{1\hskip -.37em 1}     

\begin{document}
\begin{titlepage}
\begin{centering}
 
{\ }\vspace{2cm}
 
{\Large\bf Topology Classes of Flat U(1) Bundles}\\
\vspace{0.5cm}
{\Large\bf and Diffeomorphic Covariant}\\
\vspace{0.5cm}
{\Large\bf Representations of the Heisenberg Algebra}\\
\vspace{2cm}
Jan Govaerts$^{\dag ,}$\footnote{E-mail: {\tt govaerts@fynu.ucl.ac.be}}
and Victor M. Villanueva$^{\ddag ,}$\footnote{E-mail: 
{\tt vvillanu@ifm1.ifm.umich.mx}}$^{,}$\footnote{Work done while at the
{\em Instituto de F\'{\i}sica, Universidad de Guanajuato, 
P.O. Box E-143, 37150 Le\'on, M\'exico}}\\
\vspace{0.6cm}
$^{\dag}${\em Institut de Physique Nucl\'eaire}\\
{\em Universit\'e catholique de Louvain}\\
{\em B-1348 Louvain-la-Neuve, Belgium}\\
\vspace{0.5cm}
$^{\ddag}${\em Instituto de F\'{\i}sica y Matem\'aticas}\\
{\em Universidad Michoacana de San Nicol\'as de Hidalgo}\\
{\em P.O. Box 2-82, Morelia Michoac\'an, M\'exico}\\
\vspace{2cm}
\begin{abstract}

\noindent The general construction of self-adjoint configuration space
representations of the Heisenberg algebra over an arbitrary manifold 
is considered. All such ine\-qui\-va\-lent representations 
are parametrised in terms of the topology classes of
flat U(1) bundles over the configuration space manifold. In the case of 
Riemannian manifolds, these representations are also manifestly 
diffeomorphic covariant. The general discussion, illustrated by some 
simple examples in non relativistic quantum mechanics, is of 
particular relevance to systems whose configuration space is parametrised
by curvilinear coordinates or is not simply connected, 
which thus include for instance the modular spaces 
of theories of non abelian gauge fields and gravity.

\end{abstract}

\vspace{15pt}


\end{centering} 

\vspace{45pt}


\vspace{10pt}

\noindent quant-ph/9908014\\
\noindent August 1999

\end{titlepage}

\setcounter{footnote}{0}

\section{Introduction}
\label{Sect1}

The representation theory of the Heisenberg algebra is of fundamental
importance to the ca\-no\-ni\-cal quantisation programme of Lagrangian 
and Hamiltonian systems. Indeed, through the correspondence 
principle\cite{Dirac1}, the symplectic structure of the classical 
phase space determines the algebra of (anti)commutation relations of 
quantum observables. In the case of conjugate pairs of Grassmann even 
canonical phase space variables (real under complex 
conjugation), say $q$ and $p$, the quantum observables $\hat{q}$ and $\hat{p}$
are defined by the commutation relation
\begin{displaymath}
[\hat{q},\hat{p}]=i\hbar\ \ \ ,
\end{displaymath}
as well as by the self-adjoint properties $\hat{q}^\dagger=\hat{q}$
and $\hat{p}^\dagger=\hat{p}$. This set of conditions defines 
the Heisenberg algebra in the case of a single set of canonical 
conjugate observables.

As is well known\cite{vonNeu}, when the configuration space
variable $q$ takes its values in the entire real line, there exists 
essentially only one representation of the algebra, up to unitary 
transformations, provided by the usual plane wave functions.
However, going beyond that simplest case (or trivial extensions of it
obtained by direct products) has proven difficult\footnote{Even the case
of the positive real line has been much discussed\cite{vonNeu,Trapa}.},
and it appears that no general procedure has been put forward in the case 
of an arbitrary connected manifold, or even more simply when curvilinear
coordinates are used in euclidean spaces.

In the present paper, an approach to this problem is suggested,
which, by combining algebraic and topological arguments, leads to
a general classification of the inequivalent representations of 
the Heisenberg algebra in terms of certain topological properties 
of the underlying manifold, namely those properties which may be 
characterized through a flat U(1) bundle over it. 
Moreover, when the manifold is endowed with a Riemannian metric, 
the corresponding representations are manifestly covariant under 
the diffeomorphism group of coordinate transformations, thereby including
curvilinear parametrisations of configuration space. 
Finally, the requirement of the self-adjoint properties
of the algebra, and beyond it, of the quantum dynamics,
leads to clearly identified specific restrictions 
on wave function representations.

The present discussion has no pretense to mathematical rigour, 
but rather is a physicist's constructive approach, deferring 
the more subtle issues of functional analysis to a later stage. 
As such, it also is a clear invitation to mathematical physicists 
to dwelve further into the mathematical justifications beyond 
the settings of the problem as advocated here.

The paper is organised as follows. The next section develops 
the representation theory of the Heisenberg algebra for an arbitrary discrete 
number of conjugate pairs of canonical variables, requiring 
only algebraic and topological considerations.
This is followed by Sect.~3 which ela\-bo\-ra\-tes somewhat further
on configuration and momentum space wave function representations.
Sect.~4 then applies that discussion to a more physical setting, 
at which stage geometrical and quantum dynamical aspects come into play. 
These three sections thus detail our main general results, which are
illustrated in Sects.\ref{Sect5} and \ref{Sect6} through some simple examples 
borrowed from non relativistic quantum mechanics. Finally, further 
remarks are presented in the Conclusions.

\section{The Abstract Setting: Algebra and Topology}
\label{Sect2}

Quite generally, let us consider a discrete set of real classical
conjugate observables $(q^\alpha,p_\alpha)$ ($\alpha=1,2,\dots,n$). The number
$n$ of such pairs, if infinite, could be discrete or non countable, thereby
raising further issues of mathematical justifications. For definiteness
though, we shall think of that number $n$ as being finite in the present 
discussion.

Moreover, at the present stage, the variables $q^\alpha$ and $p_\alpha$
are not necessarily a set of coordinates of the configuration space
of a physical system, and of their conjugate momenta parametrising
the cotangent bundle. Indeed, they could correspond to a collection
of conjugate observables which, at the quantum level, are assumed to
obey the Heisenberg algebra. Ne\-ver\-the\-less, since such a situation 
is certainly always encountered for the fundamental canonical phase space 
degrees of freedom of a dynamical system, and as such must thus always be 
addressed, again for definiteness we shall
think of the variables $q^\alpha$ as parametrising\footnote{Without
risk of confusion, a common notation for the coordinate parametrisation
of the manifold is used, even when its topology requires more than
one coordinate system to cover it entirely.} a specific but
otherwise arbitrary connected\footnote{Were the manifold to be disconnected,
each of its connected components would correspond to the configuration
space of a distinct physical subsystem.} differentiable manifold 
$M$ of dimension $n$. Except for its connectedness, no further properties 
of the manifold are assumed, such as compactness or the existence of 
boun\-da\-ries, or even a Riemannian metric structure compatible with its 
topology. Such additional specifications will stem from the geometrical and
physical properties of a specific system to be quantised, and are thus
considered only in Sect.\ref{Sect4}.

Let us remark that the r\^ole of the variables $q^\alpha$ is thus 
distinguished from the outset from that of their conjugate momenta $p_\alpha$, 
the $q^\alpha$ being the coordinates which parametrise the configuration space 
of a given system. The domain of values for $q^\alpha$ is thus specified 
by the choice of manifold $M$, {\em i.e.\/} of physical system,
while that of their conjugate momenta $p_\alpha$ is determined 
{\em at the quantum level\/} by the representation
theory of their Heisenberg algebra, the latter thus being defined by
the relations
\begin{equation}
[\hat{q}^\alpha,\hat{p}_\beta]=i\hbar\,\delta^\alpha_\beta\ \ ,\ \ 
\left(\hat{q}^\alpha\right)^\dagger=\hat{q}^\alpha\ \ ,\ \ 
\left(\hat{p}_\alpha\right)^\dagger=\hat{p}_\alpha\ \ ,\ \ 
\alpha,\beta=1,2,\dots,n\ \ \ .
\label{eq:Heis2}
\end{equation}
Indeed, even though the Heisenberg algebra may appear to be dual
under the combined exchange of the conjugate variables
$\hat{q}^\alpha\leftrightarrow\hat{p}_\alpha$ and a sign reversal of
$\hbar$, possibly different domains of spectral values for the operators
$\hat{q}^\alpha$ and $\hat{p}_\alpha$ render this apparent duality
in general meaningless\footnote{Except in some fortuitous cases
where they do coincide, for example when the manifold $M$ is isomorphic
to the $n$ dimensional euclidean space.}.

In order to develop an abstract representation theory of the Heisenberg
algebra, only the following two assumptions are made\footnote{A similar 
approach was applied in Ref.\cite{Gov1} to the simple case $n=1$ of a single
degree of freedom over the real line. In the course of completing this paper,
it was realised that the same point of view has already been advocated in 
Ref.\cite{DeWitt}, without however, pursuing the consequences of this 
approach to its full conclusions.}, of which the first does indeed
emphasize the distinguished r\^ole assumed in our approach by
the configuration space manifold $M$.
\begin{itemize}
\item[A1.] There exists a basis $|q>$ of the representation space which
is spanned by eigenstates of the position operators 
$\hat{q}^\alpha$ ($\alpha=1,2,\dots,n$),
whose domain of eigenvalues coincides with all the values of the
coordinates $q^\alpha$ parametrising the configuration space
manifold $M$,
\begin{equation}
\hat{q}^\alpha\,|q>\,=\,q^\alpha\,|q>\ \ \ ,\ \ \ \{q^\alpha\}\ \epsilon\ M
\ \ \ ,\ \ \ \alpha=1,2,\dots,n\ \ ;
\end{equation}
\item[A2.] The representation space of the algebra
may be endowed with an hermitian positive definite inner
product $<\cdot\,|\,\cdot>$ for which the operators $\hat{q}^\alpha$ and 
$\hat{p}_\alpha$ ($\alpha=1,2,\dots,n$) are self-adjoint.
\end{itemize}

Armed with these two assumptions, we shall now determine the necessary
conditions for such a general representation of the Heisenberg algebra
to exist. Since, given these assumptions, the knowledge of
the configuration space matrix elements of the
operators $\hat{q}^\alpha$ and $\hat{p}_\alpha$ is tantamount to
specifying the representation itself, let us
first consider the former type of matrix element, namely
$<q|\hat{q}^\alpha|q'>$. Clearly, by virtue of both assumptions,
its evaluation implies the relations
\begin{equation}
(q^\alpha - {q'}^\alpha)<q|q'>=0\ \ \ ,\ \ \ \alpha=1,2,\dots,n\ \ \ .
\end{equation}
Consequently, one necessarily has the following parametrisation
of the inner product expressed in the basis of position eigenstates
\begin{equation}
<q|q'>=\frac{1}{\sqrt{g(q)}}\,\delta^{(n)}(q-q')\ \ \ ,
\label{eq:qq}
\end{equation}
where $g(q)$ is {\em a priori\/} an arbitrary positive definite
function defined over the
configuration space manifold $M$. This result implies the
spectral decomposition of the identity operator in the position
eigenbasis $|q>$,
\begin{equation}
\one=\int_Md^nq\sqrt{g(q)}\,|q><q|\ \ \ ,
\label{eq:unit}
\end{equation}
which in turn, leads to the configuration space wave function representations 
$\psi(q)=<q|\psi>$ and $<\psi|q>=<q|\psi>^*=\psi^*(q)$ of any state $|\psi>$ 
belonging to the Heisenberg algebra representation space,
\begin{equation}
|\psi>=\int_Md^nq\sqrt{g(q)}\psi(q)|q>\ \ \ ,\ \ \ 
<\psi|=\int_Md^nq\sqrt{g(q)}\psi^*(q)<q|\ \ \ .
\label{eq:wavef}
\end{equation}
In particular, the inner product of any two states $|\psi>$ and $|\varphi>$
is given in terms of their configuration space wave functions $\psi(q)$
and $\varphi(q)$, respectively, by
\begin{equation}
<\psi|\varphi>=\int_Md^nq\sqrt{g(q)}\psi^*(q)\varphi(q)\ \ \ .
\label{eq:psiphi}
\end{equation}

Obviously, the choice of function $g(q)$ is directly related to the
absolute normalisation of the position eigenbasis states $|q>$, which is
entirely left unspecified by the assumptions A1 and A2 above. 
By choosing in (\ref{eq:qq}) a parametrisation in terms of the
positive square root of $g(q)$, the assumed positive definiteness of the
inner product is made manifest. In addition, as is made explicit in
(\ref{eq:psiphi}), when the manifold $M$ is endowed with a Riemannian
metric $g_{\alpha\beta}(q)$, the canonical choice for the function $g(q)$ 
is the determinant $({\rm det}\ g_{\alpha\beta}(q))$, thereby justifying the
form of the parametrisation used in (\ref{eq:qq}). This point will be 
elaborated on in Sect.\ref{Sect4}.

Note however that having specified the absolute normalisation of the
states $|q>$ in terms of the function $g(q)$, still does not
specify their phase, which is also left
entirely free by the assumptions A1 and A2. As will become clear
later on, this remaining freedom in the choice of phase for the position
eigenbasis plays a crucial r\^ole in the classification of inequivalent
representations of the Heisenberg algebra.

Let us now turn to the determination of the position matrix elements of 
the momentum operators $\hat{p}_\alpha$. For this purpose, consider first
the position matrix elements of the Heisenberg algebra (\ref{eq:Heis2}),
\begin{equation}
<q|\,[\hat{q}^\alpha,\hat{p}_\beta]\,|q'>=i\hbar\delta^\alpha_\beta\,
\frac{1}{\sqrt{g(q)}}\,\delta^{(n)}(q-q')\ \ \ .
\end{equation}
Since, using assumptions A1 and A2,
the l.h.s. of this identity reduces to
$(q^\alpha-{q'}^\alpha)<q|\hat{p}_\beta|q'>$, the relevant
matrix elements may be parametrised as
\begin{equation}
<q|\hat{p}_\alpha|q'>=\frac{-i\hbar}{\sqrt{g(q)}}
\frac{\partial}{\partial q^\alpha}\delta^{(n)}(q-q')\,+\,
\frac{1}{\sqrt{g(q)}}\Big[A_\alpha(q)+iB_\alpha(q)\Big]\delta^{(n)}(q-q')
\ \ \ ,
\label{eq:qpq1}
\end{equation}
where $A_\alpha(q)$ and $B_\alpha(q)$ ($\alpha=1,2,\dots,n$)
are the components of two {\em a priori\/} arbitrary real vector fields
defined over the manifold $M$.

However, these two vector fields are restricted further by the Heisenberg
algebra and the assumptions A1 and A2. Indeed, the requirement of
hermiticity of the inner product and the self-adjoint property of
$\hat{p}_\alpha$,
\begin{equation}
<q|\hat{p}_\alpha|q'>^*=<q'|\hat{p}^\dagger_\alpha|q>=
<q'|\hat{p}_\alpha|q>\ \ \ ,
\end{equation}
implies the following expression for the vector field $B_\alpha(q)$
\begin{equation}
B_\alpha(q)=-\frac{1}{2}\hbar\sqrt{g(q)}\frac{\partial}{\partial q^\alpha}
\left(\frac{1}{\sqrt{g(q)}}\right)\ \ \ .
\end{equation}
Consequently, (\ref{eq:qpq1}) reduces to
\begin{equation}
<q|\hat{p}_\alpha|q'>=\frac{-i\hbar}{g^{1/4}(q)}
\frac{\partial}{\partial q^\alpha}
\left(\frac{1}{g^{1/4}(q)}\delta^{(n)}(q-q')\right)\,+\,
\frac{1}{\sqrt{g(q)}}A_\alpha(q)\delta^{(n)}(q-q')\ \ \ .
\label{eq:qpq2}
\end{equation}

Furthermore, position matrix elements of the relations\footnote{The
relations $[\hat{q}^\alpha,\hat{q}^\beta]=0$ are obviously satisfied
for the representations considered, since the operators $\hat{q}^\alpha$
are diagonal in the position eigenbasis $|q>$ by virtue of assumption A1.} 
$[\hat{p}_\alpha,\hat{p}_\beta]=0$
may be evaluated using (\ref{eq:qpq2}), leading to the following
restrictions on the vector field $A_\alpha(q)$,
\begin{equation}
A_{\alpha\beta}(q)\equiv\,
\frac{\partial A_\beta(q)}{\partial q^\alpha}\,-\,
\frac{\partial A_\alpha(q)}{\partial q^\beta}\,=\,0\ \ ,\ \ 
\alpha,\beta=1,2,\dots,n\ \ \ .
\label{eq:Aab}
\end{equation}

Obviously, this result suggests that in fact the vector field
$A_\alpha(q)$ defines a section of a U(1) bundle over the configuration
manifold $M$, which, by virtue of (\ref{eq:Aab}), would then have to be
a flat bundle. To establish this specific characterization of the
vector field $A_\alpha(q)$, let us return to the remaining freedom in
phase redefinitions of the position eigenbasis $|q>$. Under an
arbitrary local phase transformation of these eigenvectors\footnote{Such
transformations obviously preserve the normalisation (\ref{eq:qq}).}, 
\begin{equation}
|q>_{(2)}=e^{i/\hbar\,\chi(q)}\,|q>\ \ \ ,
\end{equation}
where $\chi(q)$ is an arbitrary local scalar function over $M$,
it is straightforward to show that the vector field $A^{(2)}_\alpha(q)$
associated to the matrix elements $_{(2)}\!\!<q|\hat{p}_\alpha|q'>_{(2)}$
is then related to the vector field $A_\alpha(q)$ associated to the
matrix elements $<q|\hat{p}_\alpha|q'>$ by the transformation
\begin{equation}
A^{(2)}_\alpha(q)=A_\alpha(q)+\frac{\partial\chi(q)}{\partial q^\alpha}
\ \ \ .
\end{equation}
This result thus establishes that the vector field $A_\alpha(q)$,
which parametrises the position matrix elements of the momentum
operators $\hat{p}_\alpha$ in (\ref{eq:qpq2}), does indeed define a flat 
U(1) bundle over $M$. The existence of this bundle with connection
$A_\alpha(q)$ is thus associated to the freedom in the choice of 
phase of the position eigenbasis $|q>$, while the freedom in the 
normalisation of these states is associated to the choice of function $g(q)$.

Note that this characterization of the vector field $A_\alpha(q)$ indeed
requires that as a U(1) connection, it be globally well defined
over the entire manifold $M$. If the topology of $M$ is such 
that more than one coordinate patch
is required, position eigenstates $|q>$ associated to the overlap
of two patches could differ by a local phase transformation when passing
from one patch to the other. Such a transformation however,
then affects equally the U(1) connection $A_\alpha(q)$ in a manner
consistent with the transition functions between the coordinate
patches. Hence, the vector field $A_\alpha(q)$ is indeed a U(1)
connection which is well defined globally over $M$, as is thus also the 
corresponding configuration space representation of the Heisenberg algebra.

To conclude so far, configuration space representations of the Heisenberg
algebra over the manifold $M$ are thus characterized, on the one hand,
by the function\footnote{Which, for a Riemannian manifold,
will correspond to the determinant of the metric, hence to a geometrical
structure; see Sect.\ref{Sect4}.} $g(q)$, and on the other hand, 
by a flat U(1) bundle. However, since
arbitrary local gauge transformations within the U(1) bundle correspond to
arbitrary local phase redefinitions of the states $|q>$, and thus relate
representations of the Heisenberg algebra which are unitarily
equivalent, it is clear that all {\em inequivalent\/} representations
of the Heisenberg algebra over a manifold $M$ are classified in terms of
the topologically distinct flat U(1) bundles over that manifold,
{\em i.e.\/} the equivalence classes under {\em local\/} gauge 
transformations of U(1) gauge fields over $M$ of vanishing field strength.

In particular, a trivial flat U(1) bundle corresponds to a representation
of the algebra for which the gauge freedom of phase redefinitions of
the eigenstates $|q>$ may be used to gauge away locally the vector
field $A_\alpha(q)$ all over the manifold $M$, the field being then a pure
gauge configuration, $A_\alpha(q)=\partial\chi(q)/\partial q^\alpha$.
Otherwise, a non trivial flat U(1) bundle corresponds to a representation
of the Heisenberg algebra for which there is a topological obstruction
to the gau\-ging away of the vector field $A_\alpha(q)$ globally over the
entire manifold. Hence, the local algebraic characterization of any
representation of the Heisenberg algebra is constrained nevertheless 
by global topological properties of the configuration space manifold $M$
through the possible existence of a non trivial flat U(1) bundle over $M$
associated to that representation of the algebra.

It is well known that the topology classes of
flat U(1) bundles are characterized by their holonomies
around non trivial cycles in the manifold, or more 
precisely\footnote{The line integrals of $A_\alpha(q)$ along two
homotopically equivalent cycles differ by the integral of the field strength 
$A_{\alpha\beta}(q)$ over a connected surface in $M$ bounded by the two cycles.
This difference thus vanishes for a flat bundle.}, by the maps of the 
generators of the first homotopy group $\pi_1(M)$ into the gauge group U(1). 
If the manifold is simply
connected, only a trivial flat bundle is possible, since all holonomies
are then contractible to the identity. Hence over a simply connected manifold,
the Heisenberg algebra admits only a single representation, given by the
relations (\ref{eq:qq}) and (\ref{eq:qpq2}) with $A_\alpha(q)=0$
(or equivalently a pure gauge, {\em i.e.\/} a gradient
$A_\alpha(q)=\partial\chi(q)/\partial q^\alpha$, which simply defines
a unitarily equivalent representation).
For a non simply connected manifold however, topology classes of non trivial
flat bundles are thus characterized by their non trivial holonomies
around all non contractible cycles in configuration space.

In conclusion, unitarily inequivalent configuration space representations of
the Heisenberg algebra over an arbitrary manifold $M$ are parametrised
by two types of data. On the one hand, a function $g(q)$ which
determines the absolute normalisation of the position eigenbasis $|q>$.
On the other hand, a flat U(1) bundle over $M$ whose holonomies uniquely
characterize the representation.
The first information is of a geometrical character, since it determines 
an integration measure over the manifold for the configuration space 
representation of the inner product (see (\ref{eq:psiphi})). 
At this stage however, having not yet
bestowed the manifold with a geometry, the choice for $g(q)$ is totally
arbitrary and this information is thus not yet complete\footnote{For a 
Riemaniann manifold, a natural choice for $g(q)$ is the determinant of 
the metric.}.  The second information however, is of a global topological 
character, is complete, and is entirely specified from the local algebraic 
properties of the Heisenberg algebra representation over $M$. 

Consequently, for simply connected manifolds $M$, the Heisenberg algebra
possesses only a single configuration space representation up to
unitary transformations. This representation is characterized by
the arbitrary integration measure $d^nq\sqrt{g(q)}$ and 
some trivial flat U(1) bundle over $M$, whose section $A_\alpha(q)$
may always be taken to vanish identically. Clearly, this conclusion 
generalises to any simply connected manifold the very same fact known 
for a long time in the case of the real line\cite{vonNeu}.

In the case of non simply connected manifolds however, beyond the
trivial representation associated to a trivial flat U(1) bundle,
the Heisenberg algebra possesses an infinity of ine\-qui\-va\-lent
representations labelled by all topologically distinct non trivial flat U(1)
bundles, namely by the embeddings of the generators of the mapping class group
$\pi_1(M)$ into U(1) such that at least one of its non trivial
homotopy class generators is mapped onto a non trivial U(1) phase.

\section{Configuration and Momentum Space Wave Functions}
\label{Sect3}

Having thus completely characterized the representations of the Heisenberg
algebra over an arbitrary manifold $M$ solely in terms of algebraic
and topological considerations, let us momentarily return to some
consequences of these conclusions, beginning with the configuration space 
wave functions of states, $\psi(q)=<q|\psi>$, and the ensuing 
position and momentum operator representations. 

Given the parametrisation (\ref{eq:qpq2})
of the momentum operator configuration space
matrix elements, and the representation
(\ref{eq:unit}) of the unit operator, the configuration space
wave function representation\footnote{That of $\hat{q}^\alpha$ is of course
given by $<q|\hat{q}^\alpha|\psi>=q^\alpha\,\psi(q)$.}
of $\hat{p}_\alpha$ is provided by the differential operator
\begin{equation}
<q|\hat{p}_\alpha|\psi>=
\frac{-i\hbar}{g^{1/4}(q)}\frac{\partial}{\partial q^\alpha}
\left[g^{1/4}(q)\psi(q)\right]\,+\,A_\alpha(q)\,\psi(q)=
\frac{-i\hbar}{g^{1/4}(q)}\left[\frac{\partial}{\partial q^\alpha}+
\frac{i}{\hbar}A_\alpha(q)\right]g^{1/4}(q)\psi(q)\ \ \ ,
\label{eq:qpq3}
\end{equation}
in which one recognizes the covariant derivative for the U(1) gauge
connection $A_\alpha(q)$.
This expression is thus the appropriate generalisation of the usual
correspondence rule, which ignores the integration measure factors
$g^{1/4}(q)$---stemming from the normalisation of the position eigenbasis
$|q>$---and the possibly non trivial flat U(1) bundle 
$A_\alpha(q)$---stemming from the topological properties of the base manifold
$M$ and their possible obstruction to a complete gau\-ging 
away of a non trivial flat U(1) connection. The usual correspondence rule 
for the momentum operators in the configuration space representation is 
thus seen to be associated to the trivial representation of the Heisenberg 
algebra with $A_\alpha(q)=0$ and a trivial choice of integration measure 
$g(q)=1$. Such specific restrictions however, could prove to be incompatible 
with other properties that a particular system may be required to possess 
on physical or geometrical grounds.

Given the representation in (\ref{eq:qpq3}), it is also possible to
specify the properties\footnote{These are certainly necessary conditions,
but the issue of whether they are also sufficient shall not be addressed
specifically in the present paper, since this may be considered only
on a case by case basis.} required of the wave 
functions $\psi(q)$ for the momentum operators $\hat{p}_\alpha$ to be 
self-adjoint. A straightforward analysis thus finds that these conditions 
are expressed by the surface integrals
\begin{equation}
\int_M d^nq\,\frac{\partial}{\partial q^\alpha}
\left[\sqrt{g(q)}\ |<q|\psi>|^2\right]=
\int_Md^nq\,\frac{\partial}{\partial q^\alpha}
\left[\sqrt{g(q)}\,|\psi(q)|^2\right]\,=\,0\ \ ,\ \ 
\alpha=1,2,\dots,n\ \ .
\label{eq:bc1}
\end{equation}
Note how this set of conditions is independent of the choice
of normalisation and of local phase
of the position eigenbasis $|q>$, and thus indeed
determines the domain of {\em states\/} $|\psi>$ for which the operators
$\hat{p}_\alpha$ possess a self-adjoint representation on the
considered space of states. Nevertheless, {\em given a choice of normalisation
function\/} $g(q)$, the surface integrals (\ref{eq:bc1}) determine the 
necessary conditions defining the domain of the associated
{\em wave functions\/} $\psi(q)$ for which the {\em differential 
operators\/} (\ref{eq:qpq3}) are self-adjoint.

Another fundamental aspect of the momentum operators $\hat{p}_\alpha$
is that, being the generators of local translations on $M$, they
induce through the representation (\ref{eq:qpq2}) the parallel transport
of states along curves in $M$ in a manner consistent with the phase
transformation induced by the flat U(1) bundle. 
In particular, when taken along a closed
cycle starting and ending at a given point in $M$ of coordinates
$q^\alpha$, the corresponding position eigenstate $|q>$ is thus 
transformed back into itself, {\em up to a phase factor given by 
the holonomy of the U(1) connection $A_\alpha(q)$ along that cycle\/}. 
Hence, it is only for a trivial U(1) bundle that this holonomy is
always trivial, {\em i.e.\/} equal to unity, whatever the choice of cycle.

To make this remark more explicit, let us consider a basis of states
which diagonalises all momentum operators $\hat{p}_\alpha$. Indeed, since
all these hermitian operators commute with one another, they may be
diagonalised together in terms of states $|p>$ spanning the whole
space of states,
\begin{equation}
\hat{p}_\alpha=p_\alpha\,|p>\ \ \ ,\ \ \ p_\alpha\ \epsilon\ {\cal D}(p)\ \ \ .
\end{equation}
The range ${\cal D}(p)$ 
of spectral values $p_\alpha$ is left unspecified at this stage 
however, since it depends both on the topology of the base manifold $M$ 
as well as on the considered space of states $|\psi>$ obeying the conditions 
(\ref{eq:bc1}). The other properties of the momentum eigenbasis $|p>$ left 
undetermined are their normalisation and phase. By analogy with the 
normalisation of the position eigenbasis $|q>$ in (\ref{eq:qq}), the 
normalisation of the momentum eigenstates $|p>$
is parametrised according to\footnote{The notation used throughout 
applies when the momentum eigenvalues $p_\alpha$ form a continuous spectrum.
In the case of a discrete spectrum, the relevant expressions have to
be adapted appropriately, in an obvious manner.}
\begin{equation}
<p|p'>=\frac{1}{\sqrt{h(p)}}\,\delta^{(n)}\left(p-p'\right)\ \ \ ,
\label{eq:pp}
\end{equation}
where $h(p)$ is an arbitrary positive definite function defined over the
domain ${\cal D}(p)$ of momentum eigenvalues $p_\alpha$ 
$(\alpha=1,2,\cdots,n)$.  Clearly, this choice still leaves the phase of 
the states $|p>$ as a free degree of freedom to be specified presently.

Note also that the above choice implies the spectral decomposition
of the unit operator as
\begin{equation}
\one=\int_{{\cal D}(p)}d^np\,\sqrt{h(p)}\,|p><p|\ \ \ .
\end{equation}

The quantities of direct relevance to the change of basis which
is being considered, are the wave functions $<q|p>$. According to the 
configuration space representation (\ref{eq:qpq3}) of the momentum 
operators $\hat{p}_\alpha$, and the fact that $|p>$ is a momentum eigenstate, 
these wave functions are determined by the following set of differential 
equations\footnote{Note that these equations mean that the combination
$e^{-\frac{i}{\hbar}q^\alpha p_\alpha}g^{1/4}(q)<q|p>$ is covariantly 
constant for the covariant derivative associated to the U(1) gauge 
transformations of states.}
\begin{equation}
-i\hbar\frac{\partial}{\partial q^\alpha}
\left[\,g^{1/4}(q)\,<q|p>\,\right]\,+\,
A_\alpha(q)\,g^{1/4}(q)\,<q|p>\,=\,p_\alpha\,g^{1/4}(q)\,<q|p>\ \ \ .
\label{eq:qp1}
\end{equation}

The construction of a solution to these equations proceeds as follows.
Since they define a set of first order differential equations, they require
one integration constant, namely the wave function $<q_0|p>$ associated
to a specific point on $M$ of coordinates $q_0^\alpha$. Then, given
any such specific point $q_0$ chosen arbitrarily, consider all other points
in $M$ of coordinates $q^\alpha$, and for each of these points, an oriented 
path $P(q_0\rightarrow q)$ running from $q_0$ to $q$. In addition,
as functions of the end point $q$, this network of paths attached to 
a given $q_0$ must also define a {\sl continuous\/} set. Given such data, the 
solution to (\ref{eq:qp1}) is of the form
\begin{equation}
g^{1/4}(q)<q|p>=\Omega\left[P(q_0\rightarrow q)\right]\,
e^{\frac{i}{\hbar}(q-q_0)\cdot p}\,g^{1/4}(q_0)<q_0|p>\ \ \ ,
\end{equation}
where the notation $(q\cdot p)$ stands of course for the sum 
$(q^\alpha p_\alpha)$, and where $\Omega\left[P(q_0\rightarrow q)\right]$ 
is the path ordered U(1) holonomy along the path $P(q_0\rightarrow q)$,
\begin{equation}
\Omega\left[P(q_0\rightarrow q)\right]=
Pe^{-\frac{i}{\hbar}\int_{P(q_0\rightarrow q)}dq^\alpha A_\alpha(q)}\ \ \ .
\end{equation}
However, the normalisation condition (\ref{eq:pp}) then also requires that
\begin{equation}
\left|g^{1/4}(q_0)<q_0|p>\right|^2=\frac{1}{(2\pi\hbar)^n}\,
\frac{1}{\sqrt{h(p)}}\ \ \ ,
\end{equation}
so that necessarily
\begin{equation}
g^{1/4}(q_0)<q_0|p>=\frac{e^{i\varphi(q_0,p)}}{{(2\pi\hbar)}^{n/2}}\,
\frac{1}{h^{1/4}(p)}\ \ \ ,
\end{equation}
where $\varphi(q_0,p)$ is a specific but otherwise arbitrary real function.

Hence finally, the momentum eigenstate configuration space wave functions
are given by\footnote{It may easily be checked that the normalisation
condition (\ref{eq:qq}) is then also satisfied. Moreover, note how the
phase factor $e^{iq\cdot p/\hbar}$ constrains the spectral values of
$\hat{p}_\alpha$, in a manner depending both on the topology of the 
configuration space manifold $M$ (for example, whether it is bounded, 
periodic, or otherwise) as well as on the boundary conditions (\ref{eq:bc1}),
thereby determining the domain ${\cal D}(p)$ of momentum eigenvalues.}
\begin{equation}
<q|p>=\frac{e^{i\varphi(q_0,p)}}{{(2\pi\hbar)}^{n/2}}\,
\frac{\Omega\left[P(q_0\rightarrow q)\right]}{g^{1/4}(q)h^{1/4}(p)}\,
e^{\frac{i}{\hbar}(q-q_0)\cdot p}\ \ \ ,
\end{equation}
generalising in a transparent manner the usual plane wave solutions
of application to the trivial representation of the Heisenberg algebra
with $A_\alpha(q)=0$ and with the choices $g(q)=1$ and $h(p)=1$.
It thus appears that the arbitrary function $\varphi(q_0,p)$ is directly
related to a phase convention for the momentum eigenbasis $|p>$.
A specific choice could be such that,
\begin{equation}
e^{i\varphi(q_0,p)}\,e^{-\frac{i}{\hbar}q_0\cdot p}\,=\,1\ \ \ ,
\end{equation}
thereby absorbing into the phase of the momentum eigenbasis $|p>$
the dependency of $<q|p>$ on the point $q_0$ through
the phase function $\varphi(q_0,p)$.
Such a restriction shall not be implemented here.

Momentum wave functions $<q|p>$ are thus constructed in terms of a base
point $q_0$ and a continuous network of paths $P(q_0\rightarrow q)$ in $M$.
The possible direct and rather trivial dependency on $q_0$ through
$\varphi(q_0,p)$ has just been mentioned,
while that on the path $P(q_0\rightarrow q)$ appears
through the holonomy factor $\Omega\left[P(q_0\rightarrow q)\right]$.
In the case of a trivial flat U(1) bundle $A_\alpha(q)$, this holonomy
factor may be gauged away altogether through a phase redefinition of the
position eigenbasis $|q>$ and an appropriate redefinition of the 
function $\varphi(q_0,p)$. For a non trivial flat bundle however,
such gauging away is not possible globally over $M$, so that momentum
wave functions $<q|p>$ carry with them generally such a holonomy
phase factor dependent both on the path $P(q_0\rightarrow q)$ and the
flat U(1) connection $A_\alpha(q)$.

These remarks also indicate how the functions $<q|p>$ depend on the choices
of $q_0$ and the network of paths $P(q_0\rightarrow q)$.
Changing these choices implies thus a change in the phase of the
wave functions $<q|p>$, possibly through a redefinition both of the
function $\varphi(q_0,p)$ and the holonomies 
$\Omega\left[P(q_0\rightarrow q)\right]$. In fact, since the network of
paths is continuous, the change in holonomy would be at most
only through a phase factor common to all wave functions $<q|p>$
and associated to the holonomy of $A_\alpha(q)$ along some 
homotopically non trivial closed cycle. Since the U(1) connection is flat,
that latter holonomy in fact only depends on the homotopy class 
generator to which that
cycle belongs, and is thus a constant factor independent of $q_0$ and $p$.
Hence, any dependency on the choices of base point $q_0$ and network 
of paths $P(q_0\rightarrow q)$ may always be absorbed into a redefinition 
of the function $\varphi(q_0,p)$, namely into the phase of the momentum 
eigenbasis $|p>$, without any physical consequences.

These conclusions having been reached, it is now straightforward to
show that the momentum operators $\hat{p}_\alpha$ do indeed generate
translations in the configuration space manifold $M$, namely
\begin{equation}
Pe^{-\frac{i}{\hbar}\int_{P(q_0\rightarrow q)}\,
dq^\alpha\,\hat{p}_\alpha}\ |q_0>=
\frac{g^{1/4}(q)}{g^{1/4}(q_0)}\,
\Omega\left[P(q_0\rightarrow q)\right]\,
\Omega^{-1}\left[P(q_0\rightarrow q_0)\right]\,|q>\ \ \ .
\end{equation}
Here, $P(q_0\rightarrow q)$ is a path belonging to a continuous
network of paths with $q_0$ as base point, while $P(q_0\rightarrow q_0)$
is the specific closed path of that network connecting $q_0$ to itself,
which thus characterizes the homotopy class of the network of paths.
Hence, in the limit that the point $q$ is taken along a closed 
cycle $C(q_0\rightarrow q_0)$, the momentum operators $\hat{p}_\alpha$
indeed take the state $|q_0>$ back to itself, up to the holonomy
$\Omega\left[C(q_0\rightarrow q_0)\right]$
of the flat U(1) bundle along that cycle, as well as the U(1) mapping
$\Omega^{-1}\left[P(q_0\rightarrow q_0)\right]$
of the homotopy class of the network of paths used,
\begin{equation}
Pe^{-\frac{i}{\hbar}\int_{C(q_0\rightarrow q_0)}\,
dq^\alpha\,\hat{p}_\alpha}\ |q_0>=
\Omega\left[C(q_0\rightarrow q_0)\right]\,
\Omega^{-1}\left[P(q_0\rightarrow q_0)\right]\,|q_0>\ \ \ .
\end{equation}
When considered for all elements of the mapping class group $\pi_1(M)$,
the phases $\Omega\left[C(q_0\rightarrow q_0)\right]$ are precisely
the functions which determine the embedding of the generators
of the first homotopy group
into the gauge group U(1), namely the representation of the
Heisenberg algebra effectively being considered.

As a last application of the above developments solely based on the algebraic
and topological properties of the Heisenberg algebra representation, 
let us consider the construction of a phase space path integral 
representation of the evolution operator of a quantised system,
\begin{equation}
U(t_f,t_i)=e^{-\frac{i}{\hbar}(t_f-t_i)\hat{H}}\ \ \ .
\end{equation}
Here, $\hat{H}$ is the quantum Hamiltonian of a given system,
assumed to be self-adjoint on the con\-si\-de\-red representation space.
We choose to discuss this point here rather
than at the end of the next section, to emphasize the fact that, while
the definition of the quantum Hamiltonian requires that the configuration
manifold $M$ be endowed with a geometrical structure, the construction
of a path integral representation {\sl per se\/} only requires the
algebraic and topological considerations developed so far solely on basis
of the Heisenberg algebra and assumptions A1 and A2 above.

Given the configuration space matrix elements
$<q_f|U(t_f,t_i)|q_i>$ of the evolution ope\-ra\-tor, the steps leading to
a phase space path integral representation of that quantity are
well known (see for example Ref.\cite{Gov1}). In the present general
setting, the main differences are in the normalisation factors
$g(q)$ and $h(p)$ for position and momentum eigenstates, as well as
in the phase factors related to the U(1) holonomies appearing in
the matrix elements $<q|p>$ as constructed above. Except for these
differences, the technical details are those of the usual development,
leading in the present case to the expression
\begin{displaymath}
<q_f|U(t_f,t_i)|q_i>=\frac{\Omega\left[P(q_0\rightarrow q_f)\right]
\Omega^{-1}\left[P(q_0\rightarrow q_i)\right]}
{g^{1/4}(q_f)g^{1/4}(q_i)}\,\times
\end{displaymath}
\begin{equation}
\times\lim_{N\rightarrow\infty}\,
\int_M\prod_{i=1}^{N-1}d^nq_i\int_{{\cal D}(p)}\prod_{i=0}^{N-1}
\frac{d^np_i}{(2\pi\hbar)^n}\,
e^{\frac{i}{\hbar}\sum_{i=0}^{N-1}\epsilon\,
\left[\frac{q_{i+1}-q_i}{\epsilon}\cdot p_i\,-\,h_i\right]}\ \ \ ,
\label{eq:discrete}
\end{equation}
where,
\begin{equation}
\epsilon=\frac{t_f-t_i}{N}\ \ \ ,\ \ \ 
h_i=\frac{<p_i|\hat{H}|q_i>}{<p_i|q_i>}=h^*_i\ \ ,\ \ i=0,1,\cdots,N-1\ \ \ .
\end{equation}
Note how any dependency on the normalisation factors $g(q)$ and $h(p)$
and the holonomies $\Omega\left[P(q_0\rightarrow q)\right]$ has dropped
from the phase space integration measure defining the discretized form
of the path integral. The only dependency on these factors stems from
the external states $|q_i>$ and $|q_f>$ in a consistent manner.
As a matter of fact, apart from the overall holonomy factors
$\Omega\left[P(q_0\rightarrow q_i)\right]$ and
$\Omega\left[P(q_0\rightarrow q_f)\right]$, the only other dependency of the
path integral on the flat U(1) connection $A_\alpha(q)$ on $M$, is 
implicit through the normalised matrix elements $h_i$ of the 
quantum Hamiltonian operator.

It should be emphasized that the construct (\ref{eq:discrete})
provides an exact representation of the matrix element
$<q_f|U(t_f,t_i)|q_i>$, whose discretized form is
entirely and solely determined by the properties of the
representation of the Heisenberg algebra being used. No ambiguity whatsoever 
arises for the discretized form of the phase space path integral, 
with in particular an integration measure which is completely specified, 
and is independent of the normalisation factors $g(q)$ and $h(p)$. 
This expression is thus valid under all circumstances, whether the
configuration manifold is curved or flat, or whether it is
parametrised by curvilinear coordinates or not.
Note also that in those cases for which the integrations over the 
discretized conjugate momenta variables $p_{\alpha i}$ 
$(\alpha=1,2,\dots,n; i=0,1,\dots,N-1)$ are feasable,
the above path integral reduces to a discretized path integral
over configuration space, with the appropriate integration measure
factors for the coordinate variables $q^\alpha_i$
$(\alpha=1,2,\dots,n; i=1,2,\dots,N-1)$ obtained without any ambiguity
either, as well as a discretized expression for the Lagrangian
appearing in the exponential factor of the integrand.

\section{The Physical Setting: Geometry and Dynamics}
\label{Sect4}

Having understood how general configuration space
representations of the Heisenberg algebra
may be constructed over any arbitrary differentiable manifold $M$, in a manner
involving only algebraic and topo\-lo\-gi\-cal considerations, let us apply
these results to the dynamical description of a given physical system.
Quite generally, the description of such dynamics requires first the 
specification of a geometry on the configuration space $M$ of the system, 
in order to define a variational principle in terms of a Lagrange 
function. Typically, such a Lagrange function is of the form
\begin{equation}
L=\frac{1}{2}mg_{\alpha\beta}(q)\dot{q}^\alpha\dot{q}^\beta\,-\,V(q)
\ \ \ ,
\label{eq:Lag}
\end{equation}
the notation being self-explanatory. In particular, the functions 
$g_{\alpha\beta}(q)$ define the components of a metric on the configuration 
space $M$ of the system, expressed in the coordinate system specified by 
the variables $q^\alpha$. Note that such a description encompasses both 
general Riemannian manifolds of non vanishing scalar curvature,
as well as flat spaces parametrised by curvilinear
coordinates, such as for example the $n$ dimensional euclidean space
parametrised by spherical coordinates\footnote{Indeed, the representation
theory of the Heisenberg algebra in coordinate systems other than
cartesian is usual\-ly problematic\cite{Klauder}, and 
the approach of the previous
two sections should precisely provide the adequate framework to address
these issues, beginning with curvilinear coordinates on an euclidean 
manifold.}. In this sense, one may say that dynamics requires geometry, 
while quantisation requires algebra and topology. The purpose of this section
is to show how quantum dynamics thus results from the marriage of 
algebra, topology and geometry.

Given the Lagrange function (\ref{eq:Lag}), the canonical Hamiltonian
analysis of such a system is straightforward, leading to momenta $p_\alpha$
canonically conjugate to the coordinates $q^\alpha$---whose Poisson brackets
$\{q^\alpha,p_\beta\}=\delta^\alpha_\beta$ are thus canonical---and 
a time evolution generated by the classical Hamiltonian
\begin{equation}
H_0=\frac{1}{2m}\,p_\alpha\,g^{\alpha\beta}(q)p_\beta\,+\,V(q)\ \ \ ,
\label{eq:H0}
\end{equation}
$g^{\alpha\beta}(q)$ being the inverse of the metric $g_{\alpha\beta}(q)$.

Canonical quantisation of such a system is then defined by the Heisenberg
algebra (\ref{eq:Heis2}) associated to the configuration space $M$, thereby
leading to a space of quantum states which provides a representation of
that algebra, and thus belongs to one of the constructions described in 
Sects.\ref{Sect2} and \ref{Sect3} involving a flat U(1) bundle and its
topological characterization in terms of the first homotopy group $\pi_1(M)$.
Which of these representations is to be used for a given physical system
depends on the observed physical properties of that system\footnote{In the
same way that the choice of representations of the $su(2)$ algebra to be used 
for a space rotationally invariant quantum system for example,
depends on the physical spin content of that system.}.

Moreover, the dynamics of the quantised system derives from its
quantum Hamiltonian $\hat{H}$ which, through the correspondence principle,
should be in correspondence with the classical Hamiltonian (\ref{eq:H0}).
However, this principle does not specify uniquely the {\sl quantum\/} 
Hamiltonian, since two such operators differing only by terms of order $\hbar$
(or $\hbar^2$ for a time reversal invariant system) correspond to
identical dynamics at the classical level. In fact, in the same way as for the
Heisenberg algebra representation, the choice of a 
quantum Hamiltonian in correspondence with a classical one is also a matter
of physics, namely of the physical properties of the dynamics of the
(quantum) system being considered, the only other restriction of application 
being that the quantum Hamiltonian defines a self-adjoint operator on 
the space of states, in order to ensure a unitary time evolution.

This freedom in the choice of quantum dynamics is demonstrated by the
following two variable parametrisation of quantum Hamiltonians
$\hat{H}_{\mu,\nu}$, which are in direct correspondence with
the classical Hamiltonian $H_0$ in (\ref{eq:H0}),
\begin{equation}
\hat{H}_{\mu,\nu}=\frac{1}{2m}\
\hat{g}^{-1/4+\mu-i\nu}(\hat{q})\
\hat{p}_\alpha\ \hat{g}^{1/2-2\mu}(\hat{q})\ \hat{g}^{\alpha\beta}(\hat{q})\
\hat{p}_\beta\ \hat{g}^{-1/4+\mu+i\nu}(\hat{q})\ 
+\ \hat{V}(\hat{q})\ \ \ ,
\label{eq:Hmunu}
\end{equation}
where $\mu$ and $\nu$ are two arbitrary real variables, while
$\hat{q}^\alpha$ and $\hat{p}_\alpha$ are the self-adjoint operators whose
representations where developed in Sects.\ref{Sect2} and \ref{Sect3}.
By construction, the latter pro\-per\-ties of $\hat{q}^\alpha$ and
$\hat{p}_\alpha$ should ensure that $\hat{H}_{\mu,\nu}$ is indeed 
a self-adjoint operator on the considered representation space of 
the Heisenberg algebra. Hence, $\hat{H}_{\mu,\nu}$ is {\sl a priori\/}
a perfectly acceptable time evolution operator
for the quantised system, and only physical properties to be observed
may determine whether the association of $\hat{H}_{\mu,\nu}$ 
for specific values of $\mu$ and $\nu$ to a given physical
system is indeed appropriate. 

Note that the existence of such a two
variable parametrisation of self-adjoint quantum Hamiltonians in 
correspondence with a common classical one, is reminiscent
of von~Neumann's theory of self-adjoint extensions of differential
operators\cite{vonNeu}. In fact, there may exist some cor\-res\-pon\-den\-ce 
between the parameters $\mu$ and $\nu$ introduced here, and the von~Neumann
deficiency indices characterizing the self-adjoint extensions of the quadratic
differential ope\-ra\-tor associated to the Hamiltonian $\hat{H}_{\mu=0,\nu=0}$
as well as to the trivial configuration space representation of the Heisenberg 
algebra whose U(1) connection $A_\alpha(q)$ vanishes identically. Being beyond
the scope of the present paper, this issue shall not be pursued here.

As a matter of fact, the choice $(\mu=0,\nu=0)$ corresponds to the usual
scalar Laplacian ope\-ra\-tor over the configuration manifold, which indeed
provides for the canonical choice of quantum Hamiltonian.
In the configuration space representation, its expression is
\begin{equation}
\hat{H}_{0,0}\ \ :\ \ 
\frac{-\hbar^2}{2m}\frac{1}{\sqrt{g(q)}}\,
\left[\partial_\alpha+\frac{i}{\hbar}A_\alpha(q)\right]\,
\sqrt{g(q)}\,g^{\alpha\beta}(q)\,
\left[\partial_\beta+\frac{i}{\hbar}A_\beta(q)\right]\,+\,
V(q)\ \ \ ,
\label{eq:H00}
\end{equation}
where the ordinary derivatives 
$\partial_\alpha\equiv\partial/\partial q^\alpha$ are
replaced by U(1) covariant derivatives. Thus for example, in the case
of a system whose configuration space is the $n$ dimensional euclidean
space, and for which it is established on physics grounds
that its quantum dynamics is governed by the scalar Laplacian operator 
in cartesian coordinates, it is the Hamiltonian $\hat{H}_{0,0}$ which
is to be used for a description of the same quantum system in
any curvilinear parametrisation of its configuration space.

On the other hand, the choice $(\mu=\frac{1}{4},\nu=0)$ leads to
a quantum Hamiltonian $\hat{H}_{1/4,0}$ in direct naive correspondence
with the classical Hamiltonian (\ref{eq:H0}),
\begin{equation}
\hat{H}_{\frac{1}{4},0}=\frac{1}{2m}\hat{p}_\alpha\,
\hat{g}^{\alpha\beta}(\hat{q})\,\hat{p}_\beta\,+\,\hat{V}(\hat{q})\ \ \ .
\end{equation}
However, this operator does not produce the U(1) invariant scalar
Laplacian operator on the configuration manifold, and thus in general
determines a quantum dynamics different from that governed by $\hat{H}_{0,0}$,
even when the configuration space $M$ and its geometry $g_{\alpha\beta}(q)$
are identical, as well as the representation of the Heisenberg algebra 
which is being used, and which is characterized through the topological 
class of the flat U(1) connection $A_\alpha(q)$.

In fact, all the operators $\hat{H}_{\mu,\nu}$ differ by terms of
order $\hbar^2$, so called a ``quantum correction potential". 
An explicit calculation finds
\begin{equation}
\hat{H}_{\mu,\nu}-\hat{H}_{0,0}=\hat{\Delta}_{\mu,\nu}
\ \ \ ,
\end{equation}
with the function $\Delta_{\mu,\nu}$ given by
\begin{displaymath}
\Delta_{\mu,\nu}=\frac{\hbar^2}{2m}\,g^{\mu-1/2}\,
\partial_\alpha\left[\,g^{1/2}g^{\alpha\beta}\,\partial_\beta\,
g^{-\mu}\,\right]\ +\ \frac{\hbar^2\,\nu^2}{2m}\,g^{-2}\,g^{\alpha\beta}\,
\partial_\alpha\,g\,\partial_\beta\,g\ +
\end{displaymath}
\begin{equation}
+\ \frac{\hbar\,\nu}{2m}\,
\left[\,\left(g^{-1}g^{\alpha\beta}\,\partial_\beta\,g\right)\,p_\alpha\,+\,
p_\alpha\,\left(g^{-1}g^{\alpha\beta}\,\partial_\beta\,g\right)\,\right]\ \ \ ,
\label{eq:Deltamunu}
\end{equation}
where in this last expression,
care has to be exercised in not commuting variables which at the
quantum level correspond to non commuting operators. In general, unless
this quantity $\Delta_{\mu,\nu}$ vanishes identically, the physics implied
by the Hamiltonians $\hat{H}_{\mu,\nu}$ and $\hat{H}_{0,0}$
are clearly different.

The point made above concerning the invariance under changes of coordinates
of the quantum physics properties of a system whose $n$ dimensional euclidean 
space is parametrised by cur\-vi\-li\-near coordinates, also brings us to 
an important property of the representations of the Heisenberg algebra
constructed in Sects.\ref{Sect2} and \ref{Sect3}.
As was already indicated there, when the configuration space manifold $M$
is endowed with a Riemannian geometrical structure, which is now required 
by the dynamics, the canonical choice for the normalisation function $g(q)$
of the position eigenbasis $|q>$ is the determinant 
$g(q)={\rm det}\ g_{\alpha\beta}(q)$ of the Riemannian metric.
What makes such a choice then in fact compulsory are the relations
(\ref{eq:wavef}) and (\ref{eq:psiphi}). Indeed, these identities
establish that the configuration space wave functions $\psi(q)=<q|\psi>$
of all quantum states $|\psi>$, as well as the inner product of these
states in terms of their wave functions, are then manifestly covariant under
changes in the coordinate parametrisation of the configuration space
manifold, a fact which stems from the diffeomorphic invariant property
of the integration measure $d^nq\sqrt{{\rm det}\ g_{\alpha\beta}(q)}$
over the Riemannian manifold $M$. Hence, not only do the representations
constructed in Sects.\ref{Sect2} and \ref{Sect3} provide the most
general possible representations of the Heisenberg algebra over an
arbitrary manifold $M$, but also, and as importantly, they define 
diffeomorphic covariant representations in the case of a Riemannian
manifold. In other words, given a physical system, as well as
the as\-so\-cia\-ted representation of the Heisenberg algebra over 
its configuration space, and finally the associated quantum Hamiltonian 
$\hat{H}_{\mu,\nu}$, the description of the physical properties and time 
evolution of the system which then ensue, is entirely independent of 
the chosen coordinate parametrisation of its configuration space.

One last issue to be addressed more specifically is the self-adjoint
property required of the quantum Hamiltonian. Formally, given
the expression (\ref{eq:Hmunu}) which defines $\hat{H}_{\mu,\nu}$, 
these ope\-ra\-tors
appear to meet that condition whatever the values for the parameters
$\mu$ and $\nu$, since the operators $\hat{q}^\alpha$ and $\hat{p}_\alpha$
are assumed to be self-adjoint. Ne\-ver\-the\-less, it proves useful
to translate that requirement also in terms of the wave functions
$\psi(q,t)=<q|\psi,t>$, which are already constrained to obey the surface
term conditions (\ref{eq:bc1}). One consequence of the unitarity of the
evolution operator generated by a given quantum Hamiltonian 
$\hat{H}_{\mu,\nu}$, is the existence of a conserved probability current 
$J_\alpha(q,t)$ in the configuration space wave function representation, 
such that
\begin{equation}
\frac{\partial}{\partial t}\rho(q,t)\,+\,\frac{1}{\sqrt{g}}
\frac{\partial}{\partial q^\alpha}\sqrt{g}g^{\alpha\beta}
\,J_\beta(q,t)\,=\,0\ \ \ ,
\label{eq:current}
\end{equation}
with the probability density
\begin{equation}
\rho(q,t)=|<q|\psi,t>|^2\ \ \ .
\end{equation}
Hence, configuration space wave functions $\psi(q,t)$ are constrained 
to satisfy both the conditions (\ref{eq:bc1}) necessary for the self-adjoint
property of the momentum operators $\hat{p}_\alpha$, as well as the
probability conservation constraint (\ref{eq:current}) at all points of
the configuration space Riemannian manifold $M$. Note that
whatever the specific relationship between $J_\alpha(q,t)$ and the
wave function $\psi(q,t)$, these two classes of restrictions involve 
both the wave function itself and its first order derivatives with respect
to the coordinates $q^\alpha$.

A specific relationship between $\psi(q,t)$ and $J_\alpha(q,t)$ follows
only a choice for the quantum Hamiltonian. Given the operators 
$\hat{H}_{\mu,\nu}$ in (\ref{eq:Hmunu}), the associated probability 
conservation equation reads as in (\ref{eq:current}), with a current 
density $J_\alpha(q)$ which defines the U(1) covariant generalisation of 
the usual expression and includes a dependency on the parameters 
$\mu$ and $\nu$,
\begin{equation}
J_\alpha=-\frac{i\hbar}{2m}\,g^{-2\mu}\Bigg[\left(g^{\mu+i\nu}\psi\right)^*
\left(\partial_\alpha+\frac{i}{\hbar}A_\alpha\right)
\left(g^{\mu+i\nu}\psi\right) -
\Bigg(\left(\partial_\alpha+\frac{i}{\hbar}A_\alpha\right)
\left(g^{\mu+i\nu}\psi\right)\Bigg)^*\left(g^{\mu+i\nu}\psi\right)\Bigg]\ \ \ .
\label{eq:current2}
\end{equation}
Note that here as well, the issue---which is beyond the scope of the present
paper---of the possible relation between the variables
$\mu$ and $\nu$ and the von~Neumann deficiency indices for the self-adjoint
extensions of $\hat{H}_{0,0}$ when $A_\alpha(q)=0$, arises again
but in another disguise. Indeed, in the case of a configuration space 
with boundaries, the conservation of probability at the boundaries 
requires that the current $J_\alpha(q)$
vanishes at those points, thereby implying specific relations between
the wave function $\psi(q,t)$ and its first order derivatives
$\partial_\alpha\psi(q,t)$ at the boundaries. Typically, such relations
precisely arise also in the discussion of self-adjoint extensions of
quadratic differential operators, and are indeed in direct correspondence
with the von~Neumann deficiency indices characterizing such 
self-adjoint extensions.

\section{The Free Particle in Two Euclidean Dimensions}
\label{Sect5}

As a simple illustration of the previous general discussion,
let us consider the case of a non relativistic particle of mass $m$
free to propagate
in a two dimensional euclidean space. Such a space being simply connected,
there thus exists, up to unitary transformations,
a single re\-pre\-sen\-ta\-tion only of the associated
Heisenberg algebra, namely the trivial
one whose U(1) flat connection vanishes identically, 
$A_\alpha(q)=0$, while the
normalisation factor $g(q)$ is determined by the flat euclidean
geometry of the configuration space. However, rather than working
in cartesian coordinates, a polar parametrisation of the manifold
will be used to demonstrate that the general analysis developed
previously is perfectly adequate to consider canonical quantisation 
in curvilinear coordinates.

With the choice of parameters $(q^1,q^2)=(q^r,q^\theta)=(r,\theta)$ 
defining polar coordinates in the plane, the corresponding metric tensor is
given by
\begin{equation}
ds^2=g_{\alpha\beta}(q)dq^\alpha dq^\beta=dr^2+r^2d\theta^2\ \ \ ,
\end{equation}
with in particular $g(q)=r^2$ specifying the normalisation of the
position eigenstates $|q>$ through (\ref{eq:qq}). Consequently,
according to (\ref{eq:qpq3}), the associated configuration space
representation of the momentum operators $\hat{p}_r$ and $\hat{p}_\theta$
is
\begin{equation}
\hat{p}_r:\ \ \ -\frac{i\hbar}{\sqrt{r}}\,\partial_r\sqrt{r}
\ \ \ ,\ \ \ 
\hat{p}_\theta:\ \ \ -i\hbar\,\partial_\theta\ \ \ .
\end{equation}

Given these operators, let us turn to the choice of Hamiltonian
determining the quantum dynamics of the system, for which we shall consider
the general abstract operator $\hat{H}_{\mu,\nu}$ defined in
(\ref{eq:Hmunu}) with a vanishing potential $\hat{V}(\hat{q})$. 
In the configuration space representation being used,
this expression reduces to the following differential operator,
\begin{equation}
\hat{H}_{\mu,\nu}:\ \ -\frac{\hbar^2}{2m}
\left[\partial^2_r+\frac{(1+4i\nu)}{r}\partial_r-
\frac{4(\mu^2+\nu^2)}{r^2}+\frac{1}{r^2}\partial^2_\theta\right]\ \ \ .
\end{equation}

Determining the energy eigenstates of this operator is a straightforward
exercise. The eigenvalue spectrum is given by all real and positive values
of the energy, $E\ge 0$, as well as all positive, null and negative
integer values of the angular momentum, $\ell$, with the associated
configuration space eigenstate wave functions determined by,
\begin{equation}
\psi_{E,\ell}(r,\theta)=\left(\frac{m}{2\pi\hbar^2}\right)^{1/2}\,
e^{i\ell\theta}\,\left(r\sqrt{\frac{2mE}{\hbar^2}}\right)^{-2i\nu}\,
J_{|\alpha|}\left(r\sqrt{\frac{2mE}{\hbar^2}}\right)\ \ \ ,
\label{eq:sol1}
\end{equation}
having introduced
\begin{equation}
|\alpha|=\sqrt{\ell^2+4\mu^2}\ \ \ ,
\end{equation}
and where $J_{|\alpha|}(x)$ is the Bessel function of the first
kind of order $|\alpha|$.
These eigenstate wave functions are normalised such that,
\begin{equation}
<\psi_{E,\ell}|\psi_{E',\ell'}>=\delta_{\ell\ell'}\,\delta(E-E')\ \ \ ,
\end{equation}
with the inner product defined by (\ref{eq:psiphi}) as well as
$g(r,\theta)=r^2$.

The Bessel functions of the second kind 
$N_{|\alpha|}(r\sqrt{2mE/\hbar^2})$ do also define
solutions to the Schr\"odinger equation associated to the above
differential operator. However, such con\-fi\-gu\-ra\-tions are excluded on the
grounds that the corresponding probability density current $J_\alpha(q)$
constructed in (\ref{eq:current2}) should remain finite throughout the
entire two dimensional plane, and in particular at the origin $r=0$.

It is clear that in the limit where $(\mu,\nu)=(0,0)$, the solutions
in (\ref{eq:sol1}) reduce to the eigenfunctions of the scalar
Laplacian operator on the plane in polar coordinates. 
One among other ways to illustrate this point 
is to consider the usual normalised plane wave eigenfunctions
of the scalar Laplacian operator expressed in cartesian coordinates, namely
\begin{equation}
\frac{1}{2\pi\hbar}\,e^{\frac{i}{\hbar}(xp_x+yp_y)}\ \ \ .
\end{equation}
Introducing the parametrisation,
\begin{equation}
x=r\cos\theta\ \ ,\ \ y=r\sin\theta\ \ \ ;\ \ \ 
p_x=p\cos\varphi\ \ ,\ \ p_y=p\sin\varphi\ \ ,\ \ p=\sqrt{2mE}\ge 0\ \ \ ,
\end{equation}
these plane wave solutions are also expressed as
\begin{equation}
\frac{1}{2\pi\hbar}\,e^{\frac{i}{\hbar}(xp_x+yp_y)}=
\sum_{\ell=-\infty}^{\infty}\,\frac{1}{\sqrt{2\pi m}}\,
\left(i^{|\ell|}e^{-i\ell\varphi}\right)\,
\left(\frac{m}{2\pi\hbar^2}\right)^{1/2}\,e^{i\ell\theta}\,
J_{|\ell|}\left(r\sqrt{\frac{2mE}{\hbar^2}}\right)\ \ \ .
\end{equation}
In other words, the usual plane wave solutions associated to cartesian
coordinates and the solutions constructed in (\ref{eq:sol1}) associated to
polar coordinates, do indeed span the same space of quantum states in
the case $(\mu,\nu)=(0,0)$ which corresponds to the scalar Laplacian
on the plane as the choice for quantum Hamiltonian.

Given the above spectrum of states for the Hamiltonian $\hat{H}_{\mu,\nu}$,
it is also possible to determine the time evolution operator of the
system. The evaluation of this propagator in configuration space,
namely that of its position matrix elements 
$<r_f,\theta_f|e^{-i\Delta t\hat{H}_{\mu,\nu}/\hbar}|r_i,\theta_i>$,
may proceed either from the explicit knowledge of the eigenstates of that
operator, or from the calculation of the phase space path integral
representation in (\ref{eq:discrete}). The former approach immediately
leads to the expression\footnote{The final integration over the
momentum $p$ may be accomplished in terms of a series involving
hypergeometric functions, which is not very illuminating.}
\begin{displaymath}
<r_f,\theta_f|e^{-i\Delta t\hat{H}_{\mu,\nu}/\hbar}|r_i,\theta_i>=
\end{displaymath}
\begin{equation}
=\frac{1}{2\pi\hbar^2}\left(\frac{r_i}{r_f}\right)^{2i\nu}\,
\sum_{\ell=-\infty}^\infty\,e^{i\ell(\theta_f-\theta_i)}\,
\int_0^\infty dp\ p\,e^{-\frac{i}{\hbar}\frac{p^2}{2m}\Delta t}\,
J_{|\alpha|}\left(\frac{r_fp}{\hbar}\right)
J_{|\alpha|}\left(\frac{r_ip}{\hbar}\right)\ \ \ .
\end{equation}

Again in the particular case $(\mu,\nu)=(0,0)$, the final summation
over the angular momentum $\ell$ may explicitely be done using one of
the addition theorems for Bessel functions. One then finds,
\begin{equation}
<r_f,\theta_f|e^{-i\Delta t\hat{H}_{0,0}/\hbar}|r_i,\theta_i>=
\frac{m}{2i\pi\hbar\Delta t}\,
e^{-\frac{m}{2i\hbar\Delta t}
\left(r_f^2+r_i^2-2r_fr_i\cos(\theta_f-\theta_i)\right)}=
\frac{m}{2i\pi\hbar\Delta t}\,
e^{-\frac{m}{2i\hbar\Delta t}\left(\vec{x}_f-\vec{x}_i\right)^2}\ \ \ ,
\end{equation}
an expression which is indeed recognized as that of the propagator for
a non relativistic free particle in two dimensions, whose Hamiltonian
is simply the canonical choice $\hat{H}=\hat{\vec{p}}\,^2/(2m)$ 
associated to the scalar Laplacian on the euclidean plane.

A similar analysis starting from the evaluation of the discretized
phase space path integral (\ref{eq:discrete}) may be developed, leading
to precisely identical conclusions. However, since this calculation will be
discussed in the next section for a system which includes the free
particle as a limiting case, no further details are given here.

Nevertheless, let us emphasize that through the general approach
developed in the previous sections, a genuine canonical quantisation
of the Heisenberg algebra associated to curvilinear coordinates and
leading to physically correct results is indeed possible, as the simple
illustration of the present section has demonstrated.

\section{The Spherical Harmonic Oscillator in a Punctured Plane}
\label{Sect6}

As a second example, let us consider the two dimensional
spherical harmonic oscillator of mass $m$ and angular frequency $\omega$
oscillating in a punctured plane of which the origin $r=0$ has been
removed. The harmonic potential $V(r)=m\omega^2r^2/2$ thus serves the
purpose of an infrared re\-gu\-la\-ri\-sa\-tion 
of the solutions to the Schr\"odinger 
equation through the confinement of the particle within the potential well, 
while the removal of
the origin of the plane induces a non trivial topology of the configuration
space of this system. 

Indeed, due to the non trivial first homotopy
group $\pi_1(M)=Z\hspace{-5pt}Z$ in this case, the Heisenberg algebra admits
an infinity of unitarily inequivalent representations, labelled by
the holonomy of a flat U(1) gauge field $A_\alpha(q)$ around the
origin $r=0$. Up to U(1) gauge transformations, such flat U(1) bundles
may all be characterized by the differential 1-form
\begin{equation}
dq^\alpha\,A_\alpha(q)=\hbar\lambda d\theta\ \ \ ,
\label{eq:diffA}
\end{equation}
where $\lambda$ is an arbitrary real parameter which thus labels the
different representations of the Heisenberg algebra. In particular,
the choice $\lambda=0$ reproduces the trivial representation
which appeared in the previous section, and which is thus equivalent
to having the two dimensional plane without the puncture at $r=0$.

Consequently, working still in polar coordinates, the configuration space
representations of the momentum operators $\hat{p}_r$ and $\hat{p}_\theta$
are modified as follows,
\begin{equation}
\hat{p}_r:\ \ \ -\frac{i\hbar}{\sqrt{r}}\,\partial_r\,\sqrt{r}\ \ \ ,\ \ \ 
\hat{p}_\theta:\ \ \ -i\hbar\left(\partial_\theta+i\lambda\right)\ \ \ .
\end{equation}
Choosing the work with the general quantum Hamiltonian
$\hat{H}_{\mu\nu}$ defined in (\ref{eq:Hmunu}), the associated
differential operator then reads
\begin{equation}
\hat{H}_{\mu,\nu}:\ \ -\frac{\hbar^2}{2m}
\left[\partial^2_r+\frac{(1+4i\nu)}{r}\partial_r-
\frac{4(\mu^2+\nu^2)}{r^2}+\frac{1}{r^2}\left(\partial_\theta+i\lambda\right)^2
\right]\ +\ \frac{1}{2}m\omega^2r^2\ \ \ .
\end{equation}

Here again, it is rather straightforward to determine the
eigenspectrum of this differential operator, thereby solving the associated
Schr\"odinger equation. One then finds that the energy spectrum of the
system is discrete and given by
\begin{equation}
E_{m,\ell}=\hbar\omega\Big[\,2m+1+|\alpha|\,\Big]\ \ ,\ \ m=0,1,2,\dots\ \ ,\ \
\ell=0,\pm 1,\pm 2,\dots\ \ \ ,
\end{equation}
with the quantity $|\alpha|$ defined as
\begin{equation}
|\alpha|=\sqrt{(\ell+\lambda)^2+4\mu^2}\ \ \ ,
\label{eq:alpha2}
\end{equation}
$\ell$ being the angular momentum of the states, as previously.
The corresponding wave functions are\footnote{There exists another
class of solutions to the second order differential Schr\"odinger equation,
but those solutions are excluded on the requirement of normalisable
energy eigenstate wave functions.}
\begin{equation}
\psi_{m,\ell}(r,\theta)=\left(\frac{m\omega}{\pi\hbar}\right)^{1/2}\,
\sqrt{\frac{m!}{\Gamma(|\alpha|+m+1)}}\,e^{i\ell\theta}\,
u^{-2i\nu}\,u^{|\alpha|}\,e^{-\frac{1}{2}u^2}\,
L^{|\alpha|}_m(u^2)\ \ \ ,
\end{equation}
with
\begin{equation}
u=r\sqrt{\frac{m\omega}{\hbar}}\ \ \ ,
\end{equation}
while the $L^{|\alpha|}_m(x)$ are the usual Laguerre polynomials.
These states are normalised according to
\begin{equation}
<\psi_{m,\ell}|\psi_{m',\ell'}>=\delta_{mm'}\delta_{\ell\ell'}\ \ \ .
\end{equation}
Note how the presence of the non trivial U(1) connection parametrised
by $\lambda$ simply shifts the value of the angular momentum $\ell$
in the energy spectrum, thereby leading to a periodic spectral flow
of the energy eigenvalues. In particular, the choice $(\mu,\nu)=(0,0)$
reproduces the well known spherical harmonic oscillator spectrum
associated to the choice of the scalar Laplacian on euclidean space
in the case of the trivial Heisenberg representation, namely when $\lambda=0$.

Having diagonalised the quantum Hamiltonian of the system, it becomes
possible to compute its propagator in configuration space, in polar coordinates.
Using the complete set of energy eigenstates constructed above, a
straightforward analysis of the relevant matrix elements easily finds
\begin{displaymath}
<r_f,\theta_f|e^{-i\Delta t\hat{H}_{\mu,\nu}/\hbar}|r_i,\theta_i>=
\end{displaymath}
\begin{equation}
=\frac{m\omega}{2i\pi\hbar\sin\omega\Delta t}\,
\left(\frac{u_i}{u_f}\right)^{2i\nu}\,
e^{\frac{i}{2}\frac{\cos\omega\Delta t}{\sin\omega\Delta t}
\left(u^2_f+u^2_i\right)}\,\sum_{\ell=-\infty}^\infty\,
e^{-i\frac{\pi}{2}|\alpha|}\,e^{i\ell(\theta_f-\theta_i)}\,
J_{|\alpha|}\left(\frac{u_f u_i}{\sin\omega\Delta t}\right)\ \ \ .
\label{eq:prop1}
\end{equation}

In particular, in the case of the trivial representation of the
Heisenberg algebra, $\lambda=0$, as well as the canonical choice
of Hamiltonian $(\mu,\nu)=(0,0)$, it is possible to show that this
expression reduces to the usual and correct one for the propagator, namely,
\begin{displaymath}
<r_f,\theta_f|e^{-i\Delta t\hat{H}_{0,0}/\hbar}|r_i,\theta_i>_{|_{\lambda=0}}=
\frac{m\omega}{2i\pi\hbar\sin\omega\Delta t}\,
e^{\frac{im\omega}{2\hbar\sin\omega\Delta t}
\left[\cos\omega\Delta t(r^2_f+r^2_i)-2r_fr_i\cos(\theta_f-\theta_i)\right]}=
\end{displaymath}
\begin{equation}
=\frac{m\omega}{2i\pi\hbar\sin\omega\Delta t}\,
e^{\frac{im\omega}{2\hbar\sin\omega\Delta t}
\left[\cos\omega\Delta t(\vec{x}^2_f+\vec{x}^2_i)-
2\vec{x}_f\cdot\vec{x}_i)\right]}\ \ \ .
\end{equation}

The same expression as in (\ref{eq:prop1}) may be derived considering
the discretized phase space path integral given in (\ref{eq:discrete}).
The fact that both calculations lead to precisely the same result,
whatever the values for $\mu$, $\nu$ and $\lambda$, shows that the
approach to the canonical quantisation of the Heisenberg algebra
advocated in general terms in this paper is sound and physically consistent
in the case of an arbitrary system of curvilinear coordinates 
on its configuration space.
In particular, it is only with the specific choice of integration
measure implicit in the discretized form of the path integral 
(\ref{eq:discrete}) that the same expression as the one given in 
(\ref{eq:prop1}) may be derived for the propagator.

More explicitely, the evaluation of the path integral (\ref{eq:discrete})
proceeds as follows. A careful analysis is required first of the abstract
quantum Hamiltonian $\hat{H}_{\mu,\nu}$ in (\ref{eq:Hmunu}), in order
to determine the normalised matrix elements,
\begin{equation}
h_i=\frac{<p_i,\ell_i|\hat{H}_{\mu,\nu}|r_i,\theta_i>}
{<p_i,\ell_i|r_i,\theta_i>}\ \ \ ,\ \ \ i=0,1,\cdots,N-1\ \ \ .
\end{equation}
Here, $p_i$ stands for the eigenvalue of the radial momentum
operator $\hat{p}_r$, while the integer $\ell_i$ labels the angular
momentum value which determines the eigenvalue of the angular
momentum operator $\hat{p}_\theta$ as $\hbar(\ell_i+\lambda)$.
A straightforward calculation then finds
\begin{equation}
h_i=\frac{1}{2m}\left[p^2_i+4\hbar\nu\,\frac{p_i}{r_i}+
\frac{\hbar^2}{r^2_i}\left((\ell_i+\lambda)^2-\frac{1}{4}+
4(\mu^2+\nu^2)-2i\nu\right)\right]\,+\,\frac{1}{2}m\omega^2r^2_i\ \ \ .
\end{equation}
Hence, in the present case, the discretised phase space path integral
representation (\ref{eq:discrete}) of the propagator reads\footnote{Contrary
to the notations used in (\ref{eq:discrete}), the spectrum of the angular
momentum operator $\hat{p}_\theta$ being discrete, with eigenvalues
$\hbar(\ell+\lambda)$, the expression in (\ref{eq:discrete}) has to be
adapted appropriately in terms of summations rather than integrations
over the integers $\ell_i$.}
\begin{displaymath}
<r_f,\theta_f|e^{-i\Delta t\hat{H}_{\mu,\nu}/\hbar}|r_i,\theta_i>=
\frac{\Omega[P(q_0\rightarrow q_f)]\Omega^{-1}[P(q_0\rightarrow q_i)]}
{(r_f r_i)^{1/2}}\times
\end{displaymath}
\begin{displaymath}
\times\lim_{N\rightarrow\infty}\,
\int_0^\infty\prod_{i=1}^{N-1}dr_i\int_0^{2\pi}\prod_{i=1}^{N-1}d\theta_i
\int_{-\infty}^{+\infty}\prod_{i=0}^{N-1}\frac{dp_i}{2\pi\hbar}
\prod_{i=0}^{N-1}\left(\sum_{l_i=-\infty}^{+\infty}\frac{1}{2\pi}\right)\times
\end{displaymath}
\begin{equation}
\times e^{\frac{i}{\hbar}\sum_{i=0}^{N-1}
\left[(r_{i+1}-r_i)p_i+\hbar(\theta_{i+1}-\theta_i)(\ell_i+\lambda)-
\epsilon h_i\right]}\ \ \ .
\end{equation}

The integrations and summations over the variables $\theta_i$, $p_i$ and
$\ell_i$ are immediate, leading to the expression,
\begin{displaymath}
<r_f,\theta_f|e^{-i\Delta t\hat{H}_{\mu,\nu}/\hbar}|r_i,\theta_i>=
\frac{\Omega[P(q_0\rightarrow q_f)]\Omega^{-1}[P(q_0\rightarrow q_i)]}
{2\pi(r_f r_i)^{1/2}}\times
\end{displaymath}
\begin{displaymath}
\times\sum_{\ell=-\infty}^{+\infty}e^{i\ell(\theta_f-\theta_i)}
\lim_{N\rightarrow\infty}\,\left(\frac{m}{2i\pi\epsilon\hbar}\right)^{N/2}
\int_0^\infty\prod_{i=1}^{N-1}dr_i\ \times
\end{displaymath}
\begin{equation}
\times\prod_{i=0}^{N-1} e^{\frac{i}{\hbar}
\left[\frac{m}{2\epsilon}(r^2_{i+1}+r^2_i)-\frac{1}{2}\epsilon m\omega^2r^2_i
-\frac{\epsilon\hbar^2}{2m}\frac{1}{r^2_i}(|\alpha|^2-\frac{1}{4}-2i\nu)\right]}
e^{-\frac{i}{\hbar}\frac{m}{\epsilon}r_ir_{i+1}}
e^{-2i\nu(\frac{r_{i+1}}{r_i}-1)}\ \ \ ,
\end{equation}
where the parameter $|\alpha|$ is already defined in (\ref{eq:alpha2}).

The evaluation of the remaining integrals over the variables
$r_i$ $(i=1,2,\dots,N-1)$ now proceeds by using the techniques 
developed in\footnote{Incidentally, note that the first reference in
\cite{Peak} considered the problem of determining the correct integration
measure in a discretized form of the phase space path integral in polar
coordinates starting from the knowledge of the correct propagator in
cartesian coordinates, associated to the scalar Laplacian operator
in the euclidean plane and the trivial representation of the
Heisenberg algebra. Here, we recover in that specific instance
the same conclusions of course, but rather by working from our general
analysis of diffeomorphic covariant representations of the Heisenberg 
algebra on arbitrary manifolds, and including the possibility of non trivial
U(1) holonomies as well as the general class of quantum
Hamiltonians $\hat{H}_{\mu,\nu}$.} Ref.\cite{Peak}. After some work,
one then finds exactly,
\begin{displaymath}
<r_f,\theta_f|e^{-i\Delta t\hat{H}_{\mu,\nu}/\hbar}|r_i,\theta_i>=
\frac{m\omega}{2i\pi\hbar\sin\omega\Delta t}\,
\left(\frac{r_i}{r_f}\right)^{2i\nu}\,
\Omega[P(q_0\rightarrow q_f)]\Omega^{-1}[P(q_0\rightarrow q_i)]\times
\end{displaymath}
\begin{equation}
\times e^{\frac{im\omega}{2\hbar}\frac{\cos\omega\Delta t}{\sin\omega\Delta t}
(r^2_f+r^2_i)}\,\sum_{\ell=-\infty}^{+\infty}\,
e^{-i\frac{\pi}{2}|\alpha|}\,e^{i(\ell+\lambda)(\theta_f-\theta_i)}\,
J_{|\alpha|}\left(\frac{m\omega}{\hbar}\frac{r_fr_i}{\sin\omega\Delta t}\right)
\ \ \ .
\label{eq:prop2}
\end{equation}

Only the evaluation of the U(1) holonomies associated to the external
states is still required, and should cancel the apparent 
lack of single-valuedness in the angular dependency on $(\theta_f-\theta_i)$ 
which may seem to occur for non integer values of the U(1) holonomy $\lambda$ 
due to the factors $e^{i(\ell+\lambda)(\theta_f-\theta_i)}$ in this
expression. Given the choice (\ref{eq:diffA}), it should be clear that
we have
\begin{equation}
\Omega[P(q_0\rightarrow q_f)]\Omega^{-1}[P(q_0\rightarrow q_i)]=
e^{-i\lambda(\theta_f-\theta_i)}\ \ \ ,
\end{equation}
so that finally the result (\ref{eq:prop2}) does indeed reproduce
{\sl exactly\/}, including its absolute normalisation and phase factors,
the propagator of the system evaluated in (\ref{eq:prop1}) directly on basis
of the diagonalisation of the quantum Hamiltonian $\hat{H}_{\mu,\nu}$.

It should be emphasized that this conclusion is by no means trivial,
since the two ways to computing this propagator are entirely different.
This thus demonstrates once again the soundness of the general construction
of configuration space representations of the Heisenberg algebra on
arbitrary manifolds advocated in this paper. Note also that the
presence of the non trivial holonomy factors $\Omega[P(q_0\rightarrow q)]$
in the path integral representation (\ref{eq:discrete}) in the
case of non trivial representations of the Heisenberg algebra, is
essential to obtain the correct final expression, and to render the
propagator indeed single-valued in multiply-connected configuration space
variables such as the angle $\theta$ in the example of this section.

Before leaving this system, we should call attention to the following
issue, related to the previously pointed out spectral flow of the energy 
spectrum as a function of $\lambda$. As is well known,
in the case of the trivial representation of the Heisenberg algebra,
$\lambda=0$, as well as the choice of the scalar Laplacian operator
as quantum Hamiltonian, $(\mu,\nu)=(0,0)$, the system may easily be
solved algebraically in terms of either cartesian or helicity-like
creation and annihilation operators, the latter choice being related 
to polar coordinates and thus particularly appropriate for a system
with circular symmetry\cite{Helicity}. In particular, the ground state
Fock vacuum is annihilated by the annihilation operators of both
helicities. However, still with the choice $(\mu,\nu)=(0,0)$, as soon
as $\lambda\ne 0$, namely for non trivial representations of the
Heisenberg algebra, a Fock space construction of the spectrum of the
quantised system still seems possible---at least the quantum Hamiltonian
$\hat{H}_{0,0}$ factorizes in terms of the helicity creation and
annihilation operators---, but then the vacuum state is at
best annihilated by only one of the helicity annihilation operators,
whose helicity depends on the sign of $\lambda$. Nevertheless, 
the energy spectrum
flows with the va\-lues of $\lambda$ and with an integer periodicity of unity.
The question which this situation thus raises is whether
a Fock space construction of the quantum spectrum is possible when
$\lambda\ne 0$ (and also possibly when $(\mu,\nu)\ne(0,0)$),
and how such an approach would explain the spectral flow not only of
the energy spectrum but also of its energy eigenstates, which define
a basis of the space of states. Presumably, some type of Bogoliubov
or coherent state transformation is required, but we shall leave this
issue beyond the scope of this paper, whose primary motivation is the
presentation of the construction of representations of the Heisenberg
algebra on arbitrary manifolds.

\section{Conclusions}
\label{Sect7}

The considerations developed in this paper concerning the general
construction of configuration space representations of the
Heisenberg algebra over an arbitrary configuration manifold,
whether flat or curved, or parametrised in terms of
curvilinear coordinates, raises a series of comments and further
issues. Quantum dynamics requires the definition of a Riemannian
me\-tric structure on configuration space, whose determinant directly specifies
the normalisation of position eigenstates in order to ensure proper
covariant properties of the Heisenberg algebra representation 
under diffeomorphisms of the configuration manifold, 
namely changes of coordinate parametrisations.
Such a property is certainly a necessary requirement for a consistent 
physical description of any system. In addition, due to the local arbitrariness
in the phase of position eigenstates, a flat U(1) bundle is always
associated to any such representation of the Heisenberg algebra. In the
case of a simply connected manifold, this flat U(1) bundle may always be
trivialized globally over the entire configuration manifold, thereby
corresponding to the ordinary trivial representation of the Heisenberg
algebra. However, for configuration spaces of non trivial mapping class
group $\pi_1(M)$, an infinity of inequivalent representations becomes
possible, being labelled by the non trivial holonomies of the flat U(1)
bundle around the non contractible cycles in the configuration manifold.
Based on these general conclusions, we have also shown how to construct
the configuration space wave functions of the momentum eigenstates, as well
as representations of quantum amplitudes in terms of discretized
path integrals over phase space. Finally, through two simple examples
borrowed from non relativistic quantum mechanics, we have demonstrated
that the general approach developed here is consistent, and does
indeed possess the different features which it is advocated to achieve.

In principle, the construction developed in this paper should lead to
self-adjoint position and momentum observables, provided the necessary
restrictions on states which were considered are met. This specific issue
should thus certainly be confronted with the usual discussion of self-adjoint 
extensions of hermitian differential operators\cite{vonNeu} in terms of 
von Neumann's deficiency indices.

Especially with the second example of the spherical harmonic oscillator
on a punctured plane, it should be clear that a simple physical picture
may be developed for the non trivial holonomies associated to the
non trivial representations of the Heisenberg algebra.
Indeed, the holonomy parametrized by $\lambda$ may also be seen as
nothing else than a Aharonov-Bohm (AB) magnetic flux line\cite{AB} piercing the
plane at its origin
and in whose vector potential the harmonic oscillator is forced
to oscillate. In particular for the choice $(\mu,\nu)=(0,0)$, the solutions
constructed in this paper should reproduce, in the limit $\omega=0$,
well established results\cite{AB,AB2} for particle scattering off an infinitely 
thin AB flux line\footnote{Note how the parameter
$\omega$ may be regarded as a regularisation parameter for the non normalisable
states of a free particle. It may thus prove useful in the context of
AB scattering and/or bound state issues, to determine quantities
for finite $\omega$, and let $\omega$ vanish only in the end result.}.
Hence more generally, the non trivial holonomies associated to non trivial
representations of the Heisenberg algebra may be regarded as being due
to specific AB flux lines passing through the different holes in 
configuration space which are characterized by the first homotopy group
$\pi_1(M)$ of that space. For example, the representations of the
Heisenberg algebra in the case of a particle moving on 
a circle (see Ref.\cite{Ohnuki}) are labelled
in terms of a U(1) holonomy which may be viewed as a AB flux line
passing through the center of that circle. The same picture applies
in the case of a 2-torus with two AB flux lines, one passing through
the hole of the torus, and the other closed onto itself and
lying inside the volume of the torus.
More generally, one may imagine that for a configuration space with
a complicated maze of holes, a whole network of intertwined 
open and closed AB flux lines
threading these holes defines all the unitarily inequivalent representations
of the Heisenberg algebra associated to that configuration space, the latter
then viewed as being embedded in some higher dimensional manifold.
{\sl A priori\/}, such a picture may well apply in the case of the
modular spaces of non-abelian Yang-Mills gauge theories or theories
of gravity, the topology of these spaces being particularly rich,
and presumably at the origin of the non-perturbative dynamics in such systems.

Another question which arises is obviously the possible relationship
between the flat U(1) bundle which is a characteristic of representations
of the Heisenberg algebra, and the well known Berry phase. It would
certainly prove very clarifying to identify the connection between
these two aspects of quantised systems with multiply connected configuration
spaces.

The latter aspect also raises the issue of the possible realisation
within the general and abstract context of geometric quantisation\cite{Wood}
of the present approach to the quantisation on arbitrary manifolds.
In particular, our results are specifically reminiscent of those of
Ref.\cite{Landsman}, in which the general methods of induced representations
are applied to the quantisation on homogeneous coset spaces $G/H$ defined
by Lie groups $G$ and $H$. Indeed, in the latter case, a $H$ bundle
arises in the classification of the possible quantum realisations of
such systems.

As a matter of fact, the differences between our approach and that of
Ref.\cite{Landsman} are the following. First, our approach is applicable
to whatever curved or flat manifold, independently of any specific
symmetries that manifold may possess. In contradistinction, coset
manifolds $G/H$ possess additional Killing vectors which generate
symmetries of the manifold, and whose existence is essential to the
construction of Ref.\cite{Landsman}. Second, our approach considers
the classification of representations of the Heisenberg algebra only,
whereas that of Ref.\cite{Landsman} constructs representations of a larger
algebra which also includes the symmetry algebra induced by the Killing
vectors of the manifold (see for example Ref.\cite{Ohnuki}). 
{}From that point of view, it should be clear
that our approach could be extended to such manifolds as well, and lead
to the same conclusion that inequivalent representations would be parametrised
by some principal bundle whose base is the configuration space and whose
fiber is related to a gauge connection in the algebra of the symmetry group
of the manifold. Indeed, in the same manner as in Ref.\cite{Landsman},
configuration space wave functions would become vector valued into
some representation space of the symmetry group of the manifold,
while the r\^ole of the flat U(1) bundle in the case of the Heisenberg
algebra would be extended to some gauge connection in the relevant
symmetry group of the manifold. Hence, even though our approach is at
the same time more general (arbitrary manifold) and more particular
(representations of the Heisenberg algebra only), it seems fair to
conclude that our results and those of Ref.\cite{Landsman} are
consistent and complementary, and provide constructions of quantum
systems in curvilinear coordinates or on curved manifolds in less or
more abstract mathematical terms.

Applications of the considerations of this paper should clearly be
numerous and of varying degrees of difficulty. The case of modular spaces
of Yang-Mills gauge theories has already been mentioned, as the ultimate
example of multiply-connected configurations spaces. Far less
ambitious but yet important would the quantisation of systems whose
configuration space is a curved manifold\cite{vanNie}, or even more simply,
some flat euclidean space but parametrised by
curvilinear coordinates rather than cartesian ones. As we have seen,
even though only the trivial representation of the Heisenberg algebra
is then relevant in the latter instance---the mapping class group of such 
manifolds being trivial---, nevertheless a correct treatment of
the factor $g(q)={\rm det}\,g_{\alpha\beta}(q)$
determined from the metric tensor is essential in obtaining the
correct canonical quantisation of the system in curvilinear coordinates,
and in particular in defining self-adjoint representations of the
position and momentum operators on the configuration manifold.
One class of systems where this latter point may be particularly
relevant is the parametrisation of the $N$-body problem in terms of
hyperspherical coordinates\cite{Chapuisat}.

\section{Acknowledgments}

Profs. J.C. Alvarez, J.-P. Antoine, J.L. Lucio-Martinez
and A. Magnus are acknow\-led\-ged for useful discussions.
This work has been partially supported by the grant
CONACyT 3979P-E9608 under the terms of an agreement
between the CONACyT (Mexico) and the FNRS (Belgium).

\end{document}